\documentclass[times,twocolumn]{aastex7}
\hypersetup{urlcolor=magenta}

\usepackage{CJK}
\usepackage{enumitem}
\usepackage{amsmath}
\usepackage{multirow}
\usepackage{color}
\usepackage{bm}
\usepackage{graphicx}
\shorttitle{Pulsation Mode Switching in Two BLAPs}

\shortauthors{MA et al.}

\begin{document}
\begin{CJK*}{UTF8}{gbsn}
\renewcommand{\textsc}[1]{#1}

\title{Mode Switching in Two Blue Large Amplitude Pulsators observed by OGLE, KMTNet and DREAMS}
\author[0009-0006-8010-4927]{Hao Ma (马皓)}
\affiliation{Department of Astronomy, Westlake University, Hangzhou 310030, Zhejiang Province, China}
\email{mahao@westlake.edu.cn}

\author[0000-0001-8317-2788]{Shude Mao (毛淑德)}
\affiliation{Department of Astronomy, Westlake University, Hangzhou 310030, Zhejiang Province, China}
\email[show]{shude.mao@westlake.edu.cn}

\author[0000-0003-0562-5643]{Seung-Lee Kim}
\affiliation{Korea Astronomy and Space Science Institute, Daejeon 34055, Republic of Korea}
\email{slkim@kasi.re.kr}

\author[0000-0003-0626-8465]{Hongjing Yang (杨弘靖)}
\affiliation{Department of Astronomy, Westlake University, Hangzhou 310030, Zhejiang Province, China}
\email{yanghongjing@westlake.edu.cn}

\author[0000-0001-6000-3463]{Weicheng Zang (臧伟呈)}
\affiliation{Department of Astronomy, Westlake University, Hangzhou 310030, Zhejiang Province, China}
\email{zangweicheng@westlake.edu.cn}

\author[0000-0002-4503-9705]{Tianjun~Gan (干天君)}
\affiliation{Instituto de Astrof\'{i}sica de Canarias (IAC), E-38205 La Laguna, Tenerife, Spain}
\affiliation{Departamento de Astrof\'isica, Universidad de La Laguna (ULL), E-38206 La Laguna, Tenerife, Spain}
\email{tjgan@foxmail.com}

\author[0000-0002-0856-3663]{Steve Heathcote}
\affiliation{Cerro Tololo Inter-American Observatory/NSF NOIRLab, Casilla 603, La Serena, Chile}
\email{steve.heathcote@noirlab.edu}

\author[0000-0003-0043-3925]{Chung-Uk Lee}
\affiliation{Korea Astronomy and Space Science Institute, Daejeon 34055, Republic of Korea}
\email{leecu@kasi.re.kr}

\author[0000-0001-6832-4325]{Tao Wu}  
\affiliation{Yunnan Observatories, Chinese Academy of Sciences, Kunming 650216, People’s Republic of China}
\affiliation{International Center of Supernovae at the Yunnan Key Laboratory, Kunming 650216, People’s Republic of China}
\affiliation{Center for Astronomical Mega-Science, Chinese Academy of Sciences, Beijing 100012, People’s Republic of China}
\affiliation{University of Chinese Academy of Sciences, Beijing 100049, People's Republic of China}
\email{wutao@ynao.ac.cn}

\author{Dong-Jin Kim}
\affiliation{Korea Astronomy and Space Science Institute, Daejeon 34055, Republic of Korea}
\email{keaton03@kasi.re.kr}

\author[0009-0000-0099-5325]{Zhenhao Wang (王圳豪)}
\affiliation{Department of Astronomy, Westlake University, Hangzhou 310030, Zhejiang Province, China}
\affiliation{Department of Astronomy, Tsinghua University, Beijing 100084, China}
\email{wangzhenhao@westlake.edu.cn}

\author[0009-0003-7681-3702]{Yaosong Yu (于耀淞)}
\affiliation{Department of Astronomy, Westlake University, Hangzhou 310030, Zhejiang Province, China}
\email{ayuyaosong@gmail.com}

\author[0000-0003-4625-8595]{Qiyue Qian}
\affiliation{Department of Astronomy, Westlake University, Hangzhou 310030, Zhejiang Province, China}
\affiliation{Department of Astronomy, Tsinghua University, Beijing 100084, China}
\email{qqy22@mails.tsinghua.edu.cn}

\author[0000-0001-5651-9440]{Yuchen Tang (唐雨辰)}
\affiliation{Department of Astronomy, Westlake University, Hangzhou 310030, Zhejiang Province, China}
\email{tangyuchen@westlake.edu.cn}

\author[0009-0005-0410-8451]{Yuxin Shang (尚钰欣)}
\affiliation{Department of Astronomy, Tsinghua University, Beijing 100084, China}
\email{shangyx22@mails.tsinghua.edu.cn}

\author[0000-0002-1287-6064]{Zhixing Li (李知行)}
\affiliation{Department of Astronomy, Westlake University, Hangzhou 310030, Zhejiang Province, China}
\email{lizhixing@westlake.edu.cn}

\author{Tomas Ahumada}
\affiliation{Cerro Tololo Inter-American Observatory/NSF NOIRLab, Casilla 603, La Serena, Chile}
\email{tomas.ahumada@noirlab.edu}

\author[0000-0003-1587-3931]{Timothy Abbott}
\affiliation{Cerro Tololo Inter-American Observatory/NSF NOIRLab, Casilla 603, La Serena, Chile}
\email{tim.abbott@noirlab.edu}

\author[0000-0002-2651-7038]{Guillermo Damke}
\affiliation{Cerro Tololo Inter-American Observatory/NSF NOIRLab, Casilla 603, La Serena, Chile}
\email{guillermo.damke@noirlab.edu}

\author[0000-0003-4432-5037]{Konstantina Boutsia}
\affiliation{Cerro Tololo Inter-American Observatory/NSF NOIRLab, Casilla 603, La Serena, Chile}
\email{konstantina.boutsia@noirlab.edu}

\author[0000-0001-6455-9135]{Alfredo Zenteno}
\affiliation{Cerro Tololo Inter-American Observatory/NSF NOIRLab, Casilla 603, La Serena, Chile}
\email{alfredo.zenteno@noirlab.edu}

\author[0000-0001-7016-1692]{Przemek Mr\'{o}z}
\affiliation{Astronomical Observatory, University of Warsaw, Al. Ujazdowskie 4, 00-478 Warszawa, Poland}
\email{pmroz@astrouw.edu.pl}

\author[0009-0007-0032-4098]{Andong Xu (徐安东)}
\affiliation{Department of Astronomy, Westlake University, Hangzhou 310030, Zhejiang Province, China}
\email{xuandong@westlake.edu.cn}

\correspondingauthor{Shude Mao}

\begin{abstract}
Blue large-amplitude pulsators (BLAPs) are rare, hot, short-period pulsating stars whose rapid oscillations provide a unique probe of stellar interiors and atmospheres. In this paper, we study two short-period BLAPs in the Galactic bulge, OGLE-BLAP-142 ($P \approx 6.6$~min) and OGLE-BLAP-148 ($P \approx 7.5$~min), by combining light curves from the OGLE, KMTNet, and DREAMS surveys, spanning 17 years with cadences ranging from minutes to hours to days. We perform Generalized Lomb-Scargle periodogram analyses of their light curves and identify, for each BLAP, three closely spaced pulsation modes separated by $\sim 1\%$. The dominant mode can switch within timescales of order a year, accompanied by a likely blueward color shift ($\sim 0.1$--$0.2$ mag).
The yearly period excursions of the resolved mode families imply apparent $|\dot{P}/P|$ values of order $10^{-6}$--$10^{-4}\,\mathrm{yr^{-1}}$, large enough to mimic or exceed smooth evolutionary period changes. One mode family in OGLE-BLAP-148 shows a coherent decade-long drift at the $\sim10^{-5}\,\mathrm{yr^{-1}}$ level, but its failure to extrapolate to the earliest and latest epochs suggests a transient mode-family drift rather than secular evolution. 
Such close-spaced multiple pulsations and mode switching may be common among BLAPs with periods shorter than 10 min, and their physical origin remains unclear.

\end{abstract}


\section{Introduction}\label{sec:intro}

Blue Large-Amplitude Pulsators (BLAPs) are a recently discovered class of hot pulsating stars characterized by short periods, large photometric amplitudes, high effective temperatures, and unusual evolutionary states. The class was first identified by the Optical Gravitational Lensing Experiment (OGLE, \citealt{OGLEIV}), which discovered 14 objects exhibiting large-amplitude variability with periods of roughly 20--40 minutes \citep{2017NatAs...1E.166P}. Their light curves resemble those of classical radial pulsators such as RR Lyrae stars or Cepheids, but on dramatically shorter timescales and at much higher temperatures ($\sim$ few $\times 10^{4}$K).

Since their discovery, the number of known BLAPs has increased rapidly owing to the growth of large-scale time-domain surveys, including OGLE, the Zwicky Transient Facility (ZTF, \citealt{Kupfer2019_HighGravityBLAPs}), OmegaWhite (\citealt{Ramsay2018_GaiaDR2BLAPs}), TMTS \citep{Lin2023_TMTSBLAP1}, SkyMapper (\citealt{Chang2024_SkyMapperBLAP}) and combinations \citep{McWhirterLam2022_GaiaZTFBLAPs}. The known population has expanded from the original 14 objects to nearly 200 by 2025 \citep{Pietrukowicz2025_ObservationalParameters,2025AcA....75..223B}. These surveys have revealed a broader diversity of periods, amplitudes, and spectroscopic properties than initially recognized, including possible subclasses such as high-gravity BLAPs \citep{Kupfer2019_HighGravityBLAPs}.

Observationally, BLAPs typically exhibit pulsation periods of a few minutes to about one hour, amplitudes of $\sim 0.05$--$0.4$ mag in the optical, effective temperatures of $\sim 25,000$--$40,000$ K, and surface gravities around $\log g \sim 4$--$5.5$ \citep{2017NatAs...1E.166P,ByrneJeffery2018_RadiativeLevitationBLAPs, Kupfer2019_HighGravityBLAPs, Kim2025_KMTNetBLAPs}. Spectroscopic observations often indicate helium enhancement and unusual atmospheric abundances, while several systems exhibit significant radial velocity variability suggestive of binary evolution \citep{Pigulski2022_HD133729,2025AcA....75..223B, Lin2023_TMTSBLAP1, Kim2025_KMTNetBLAPs}. Measurable period changes, ${\dot P}/P \sim (10^{-7} - 10^{-5})\,{\rm yr}^{-1}$, have also been detected in several objects \citep{2025AcA....75..223B}, implying secular stellar evolution on $10^5-10^7$\,yr timescales.

BLAPs are generally interpreted as evolved, stripped, or otherwise binary-processed hot stars, although their precise formation channels remain uncertain. Proposed evolutionary pathways include low-mass helium-core pre-white dwarfs with residual hydrogen-shell burning \citep{Romero2018_EvolutionaryStatusBLAPs}, core-helium-burning stars \citep{WuLi2018_EvolutionaryStatusBLAPs}, post-common-envelope remnants affected by radiative levitation \citep{ByrneJeffery2018_RadiativeLevitationBLAPs}, shell-helium-burning hot subdwarfs \citep{Xiong2022_ShellHeBurningBLAPs}, and merger products \citep{KolaczekSzymanski2024_LowMassWDMergerBLAPs,Zhang2023_WDMSMergerBLAPs}. Their observed rarity---almost 200 objects among more than $4\times 10^8$ surveyed stars \citep{2025AcA....75..223B}---is consistent with the extremely short-lived nature of the relevant evolutionary phases.

The pulsations are usually attributed to radial modes driven by the $\kappa$-mechanism associated with the iron-group opacity bump. Radiative levitation of iron-group elements, especially Fe and Ni, appears to play a critical role in producing the instability required for pulsation driving \citep{ByrneJeffery2018_RadiativeLevitationBLAPs,ByrneJeffery2020_FaintBlueStars,Jeffery2025_NonlinearBLAPModels}. Most models favor fundamental radial pulsations, although overtone solutions may exist in some systems.

Some BLAPs are already known to have pulsation more complicated than a single stable radial mode. The final OGLE search in the inner Galactic bulge reported additional periodicities in several BLAPs after prewhitening the dominant signal and its aliases \citep{2025AcA....75..223B}, while \citet{Kim2025_KMTNetBLAPs} identified OGLE-BLAP-006 as a multimode pulsator and suggested possible switching among closely spaced frequencies. These studies establish that multi-periodicity occurs in BLAPs, but most previous detections were based on time-averaged spectra, residual peaks, or close-frequency structures. It therefore remains unclear whether the additional frequencies are persistent independent modes, beating or modulation products, aliases, or transient residuals caused by timing variations. More importantly, the mechanism that selects the dominant observable mode when several frequencies are present remains unknown.

Such behavior could be important because an exchange of dominance between closely spaced pulsation families can mimic, or even exceed, the apparent period changes expected from smooth stellar evolution. In this work, we combine photometry from OGLE, the Korea Microlensing Telescope Network (KMTNet; \citealt{2016JKAS...49...37K}), and the DECam Rogue Earths and Mars Survey (DREAMS; \citealt{Yang2026}) to examine two short-period BLAPs in detail: OGLE-BLAP-142 and OGLE-BLAP-148. We report the first clear epoch-resolved cases in which multiple resolved, percent-separated BLAP period families are seen to exchange dominance over year timescales. In both objects, the additional periodicities are not merely low-amplitude residual peaks in a combined periodogram; they recur across independent observing seasons and can become the dominant signal in the light curve.

This paper is organized as follows. In \S\ref{sec:dreams}, we describe the OGLE, KMTNet, and DREAMS data sets and the recovery of known OGLE BLAPs in the DREAMS footprint. In \S\ref{sec:recovery}, we present the period determinations and mode-switching behavior of OGLE-BLAP-142 and OGLE-BLAP-148. We summarize our results and discuss the implications for multi-mode BLAP pulsation in \S\ref{sec:discussion}.

\section{OGLE, KMTNet, and DREAMS Data \label{sec:dreams}}

Our initial goal was to search for new BLAP candidates using the DREAMS data. However, in the process, we recovered all 14 BLAPs previously identified by OGLE within our five-square-degree footprint \citep{2017NatAs...1E.166P,2023AcA....73....1B,2025AcA....75..223B}: OGLE-BLAP-006, 019, 126, 133, 138, 140, 142, 143, 145, 146, 147, 148, 151, and 153. Among these, OGLE-BLAP-142 and OGLE-BLAP-148 show the clearest changes in their dominant pulsation modes when combining the OGLE, DREAMS, and KMTNet data. We therefore focus on these two objects in this work.

OGLE-BLAP-142 is located at equatorial coordinates (J2000) $(\alpha, \delta) = (17{:}54{:}31.91, -30{:}07{:}44.4)$ and Galactic coordinates $(\ell, b) = (-0.0298, -2.2758)$; its mean magnitudes are $\langle I\rangle=18.968$ and $\langle V\rangle=20.263$ mag from OGLE. OGLE-BLAP-148 is located at $(\alpha, \delta) = (17{:}55{:}58.02, -29{:}42{:}35.4)$ and $(\ell, b) = (0.4893, -2.3335)$; its mean magnitudes are $\langle I\rangle=19.498$ and $\langle V\rangle=20.401$ mag \citep{2025AcA....75..223B}.

The OGLE data were obtained with the 1.3~m Warsaw Telescope at Las Campanas Observatory in Chile. KMTNet observations were carried out using three 1.6~m telescopes, each equipped with a $4~{\rm deg}^{2}$ camera, located in Chile, South Africa, and Australia. We collect OGLE-IV data spanning 2010 to 2024 presented by \cite{2025AcA....75..223B} and KMTNet data from 2016 to 2024. Since 2016, OGLE reduced its observing cadence in the fields containing these two BLAPs, from about 10--30 exposures per night to about 3--10 exposures per night. In the same year, KMTNet began its regular survey, with a cadence of $\Gamma = 2~{\rm hr}^{-1}$ for OGLE-BLAP-142 and $\Gamma = 4~{\rm hr}^{-1}$ OGLE-BLAP-148, resulting in 5--10 times more KMTNet data than OGLE data in the overlapping period. Therefore, in our analysis we adopt OGLE data from 2010 to 2015 and KMTNet data from 2016 to 2024 since the post-2015 OGLE seasons contain relatively few data points to yield reliable period measurements. One special season is 2020. Due to COVID-19, the KMTNet Chile and South Africa sites were closed for most of the year, and the total number of observations was reduced to roughly one third of a typical season, leading to weaker constraints on the BLAP properties in that year.

The OGLE time-series photometry analyzed here is in the $I$ band. KMTNet also observed predominantly in $I$; among the usable 2016--2024 KMTNet exposures retained in our light curves, the $V$ band only accounts for $\sim 9\%$ for OGLE-BLAP-142 and OGLE-BLAP-148. We use the $I$-band data for the period analysis and use the more sparsely sampled KMTNet \(V\)-band data to provide insight into possible color evolution.

In both surveys, most images were taken in the $I$ band, with a fraction of exposures in the $V$ band to provide color information. We use the $I$-band data for the period analysis and the KMTNet $V$-band data to investigate the color evolution of the BLAPs.

The KMTNet images are processed using the updated \texttt{pySIS} package \citep{pysis, Yang2024_pysis5_RAMP1, Yang2025_RAMP2}, which performs difference image analysis. Because the distortion varies significantly over the 9-year baseline, the $I$-band images were reduced in groups of two or three seasons: 2016--2017, 2018--2020, 2021--2022, and 2023--2024. A common master reference image was used to construct the reference frame for each group, ensuring a consistent flux scale across all seasons. For the $V$-band images, the higher noise level dominates over the systematic effects of distortion, so we processed the full 9-year data set as a single group. For each group, a baseline flux of the target was measured on the reference image. The total flux in each exposure was then obtained by adding this baseline flux to the corresponding difference flux. The total instrumental fluxes, in digital units, were converted to magnitudes using the KMTNet empirical zero-point of $28.0$ mag. Because both targets are faint and suffer from significant blending, the measured baseline flux, and therefore the magnitude zero-point, might be inaccurate. However, this inaccuracy introduces only a constant offset in the light curves across all seasons and consequently does not affect the period and color variations analysis (Section~\ref{sec:modes}), both of which rely on relative flux changes.

We also include data from the DREAMS program, which uses the Dark Energy Camera (DECam; \citealt{DECam2008,DECam2015}) mounted on the 4m Blanco telescope in Chile. Since the 2025 season, DREAMS has monitored a $\sim 5~\mathrm{deg}^2$ bulge field with minute-level cadence. Approximately 80\% of the observations were obtained in the $z$ band, with the remaining in the $r$ band. In this work, we use only the $z$-band data. The 2025 DREAMS data were collected from DREAMS Data Release I \citep{Yang2026}, comprising 883 data points for OGLE-BLAP-142 and 890 data points for OGLE-BLAP-148, corresponding to roughly 20 hours of observations. The 2026 season is still ongoing, and the DREAMS team provided data obtained between February 23 and June 20, primarily through Director's Discretionary Time observations, yielding 1948 exposures for OGLE-BLAP-142 and 1966 exposures for OGLE-BLAP-148.

We converted the native DREAMS timestamps, recorded as Modified Julian Dates at CTIO, and KMTNet, recorded as Heliocentric Julian Date, to ${\rm BJD_{TDB}}$ following \cite{2025AcA....75..223B}, placing the DREAMS, KMTNet and OGLE time axes on the same system \citep{2010PASP..122..935E}.

Figures~\ref{fig:lc142100bin} and \ref{fig:lc148100bin} in Appendix \S\ref{sec:folded-lightcurves} display the phase-folded light curves of the two BLAPs from 2010 to 2026. Overall, for OGLE-BLAP-142, the $\sim$20 hr of DREAMS data provides a constraining power on the light curves comparable to that of the $\sim$150 hr of KMTNet data per season, roughly consistent with the total number of collected photons in each survey. The period and amplitude data use 100\,s OGLE $I$, 60\,s KMTNet $I$, and 42 or 60\,s DREAMS $z$ exposures. The typical target cadence is $\sim1.2$--$1.5$ min within DREAMS observing blocks, $\sim10$--$30$ min for KMTNet, and $\sim20$ min for OGLE. We verified that finite-exposure phase smearing attenuates the measured semi-amplitudes slightly but does not shift the pulsation periods (Appendix~\S\ref{sec:observing_setup}). In addition, the constraining power of the OGLE data is generally lower than that of KMTNet and the 2025 DREAMS data, except for the 2020 KMTNet season.

\section{Period Determinations and Mode Switching in Two BLAPs \label{sec:recovery}}

For period determination, we used Generalized Lomb--Scargle periodograms (GLS, \citealt{1976Ap&SS..39..447L,1982ApJ...263..835S,2009A&A...496..577Z}) on the corrected time axis and examined OGLE, KMTNet, and DREAMS data season by season to identify persistent frequency families and their annual behavior.

\subsection{OGLE-BLAP-142}

\begin{figure*}
\centering
\includegraphics[width=0.95\linewidth]{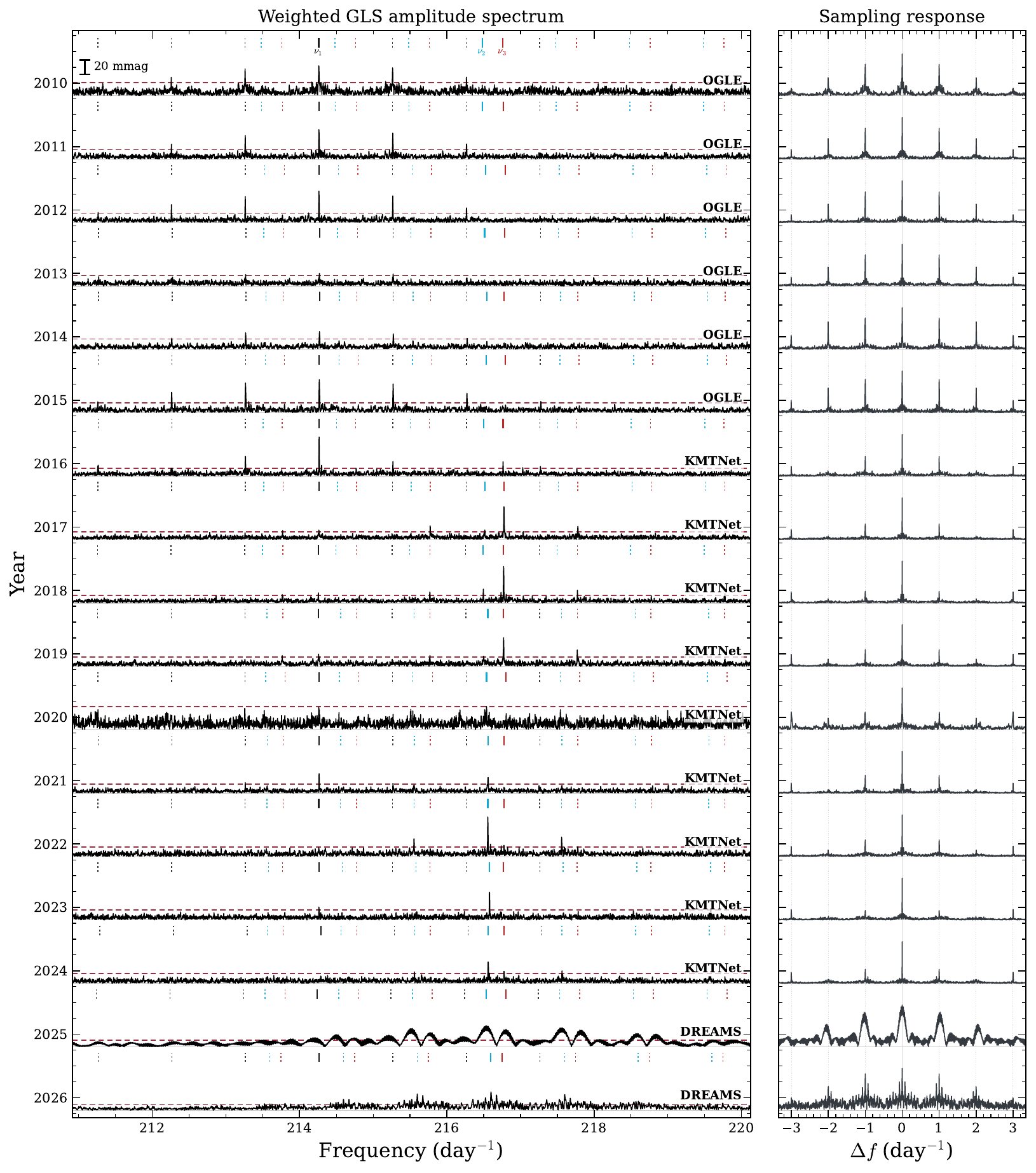}
\caption{Seasonally weighted GLS semi-amplitude spectra (left) and timestamp sampling responses (right) for OGLE-BLAP-142. The horizontal dark-red dashed lines are empirical 1\% FAP noise thresholds. The black, cyan and red markers denote the maxima of three frequencies, \((\nu_1,\nu_2,\nu_3)=(214.266,216.556,216.775)~{\rm day}^{-1}\), corresponding to \((P_1,P_2,P_3)=(6.7206,6.6495,6.6428)~{\rm min}\). Solid lines mark the adopted frequencies, and dashed lines mark their $\pm 1, \pm 2,  ...$ day aliases. The dominant family changes between 2016 and 2017, and again between 2019 and 2021. The 2020 season is less reliable because only the Australia site operated during the COVID-related interruption.}
\label{fig:gls142}
\end{figure*}

Figure~\ref{fig:gls142} shows seasonally weighted GLS semi-amplitude spectra in mmag. The dark-red dashed lines indicated the empirical 1\% False Alarm Probability (FAP) noise thresholds. To determine this threshold, we fitted and subtracted the three locally optimized frequency families and their first harmonics. We then generated 1000 fixed-timestamp residual-bootstrap realizations, and adopted the 99th percentile of the largest semi-amplitude in the displayed frequency interval.

The right-hand panel shows the timestamp sampling response (sampling-window function)
$$W(\Delta f)=\frac{\left|\sum_i w_i\exp\!\left(2\pi\mathrm{i}\,\Delta f\,t_i\right)\right|}{\sum_i w_i},\qquad w_i=\sigma_i^{-2},$$
where \(t_i\) and \(\sigma_i\) are the observation time and photometric uncertainty, \(w_i\) is the inverse-variance weight, and \(\Delta f\) is a frequency offset. The denominator is adopted such that \(W(0)=1\). Because \(W\) contains no flux or residual values, its peaks directly locate the sampling aliases at \(f+\Delta f\) of a signal at \(f\). Thus, for a particular signal at \(f\), \(W(\Delta f)\) ensures the pattern of peaks that the actual sampling would produce at \(f+\Delta f\).

The candidate frequencies lie above the nominal median-cadence Nyquist values for OGLE and KMTNet, but irregular sampling defines no unique classical Nyquist frequency \citep{VanderPlas2018_LombScargle}. Appendix~\S\ref{sec:sampling_validation} provides more detailed tests about their recovery and aliasing directly with the actual time series, and demonstrates our results are robust.

OGLE-BLAP-142 shows at least three recurrent frequency families (Fig.~\ref{fig:gls142}). The first family, $P_1\approx6.7206$ min, dominates the early OGLE data and the 2016 KMTNet season. During 2017--2019, the dominant power shifts to a shorter-period family near $P_3 \approx 6.6428$ min. The 2020 season is poorly constrained because of limited site coverage, but for 2020--2021 the modes are mixed. From 2022 onward, the dominant KMTNet power is associated with a third family near $P_2 \approx 6.6495$ min, the same family recovered by DREAMS in 2025 and 2026. Notice that for KMTNet, the periods are sharply peaked while those for DREAMS are broad due to the short duration of its observation window.

OGLE-BLAP-142 therefore appears to switch between three discrete frequency families rather than following a single (drifting) period. The light curves folded according to the dominant period are shown in Fig.~\ref{fig:lc142100bin}. Over 17 years, the light curve shape, zero-point and amplitude visibly changed: the amplitude range from 0.08 mag to 0.21 mag. In addition, some light curves are symmetric (e.g., for 2017), while others are more see-saw-like (e.g., for 2012).

\subsection{OGLE-BLAP-148}

\begin{figure*}
\centering
\includegraphics[width=\linewidth]{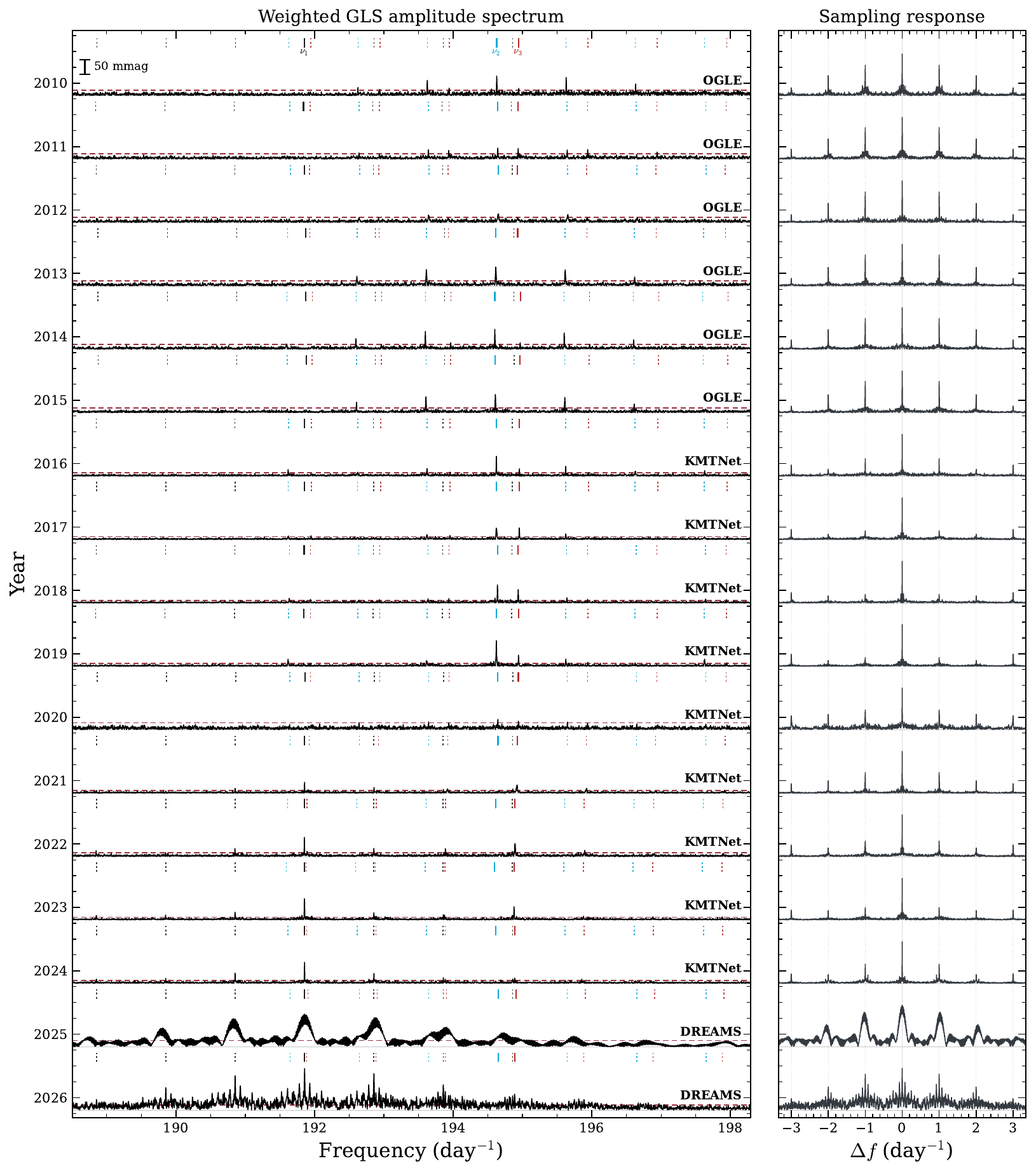}
\caption{Seasonally
weighted GLS semi-amplitude spectra (left) and timestamp sampling responses (right) for OGLE-BLAP-148.} The three frequencies are \((\nu_1,\nu_2,\nu_3)=(191.855,194.623,194.943)~{\rm day}^{-1}\) corresponding to periods \((P_1,P_2,P_3)=(7.5057,7.3989,7.3868)~{\rm min}\). The dominant peak changes between 2019 and 2021, while the subdominate component ($P_3$) remains visible through most seasons.
\label{fig:gls148}
\end{figure*}

Figure~\ref{fig:gls148} was constructed using the same procedures for calculating the amplitude spectra, FAP thresholds, and sampling windows as Fig.~\ref{fig:gls142}.

The earliest dominant OGLE period is near $P_2 \approx 7.3989$ min, while the latest DREAMS data recover a dominant family near $P_1 \approx7.5057$ min (corresponding to $\nu_2, \nu_1 \approx 194.623, 191.855~{\rm day}^{-1}$, respectively). The two periods differ by $\Delta P/P\approx1.44\%$. In addition, a secondary peak around $P_3 \approx 7.3868$ min is present throughout the whole 17 year span but with different amplitudes. 

The KMTNet periodograms fill the gap between OGLE and DREAMS. The $P_2$ family remains dominant through 2020. From 2021 to 2024 the dominant peak follows the $P_1$ family while the $P_2$ family weakens. The $P_3$ family is present throughout the KMTNet data (with some frequency drift over a decade period, see Fig. \ref{fig:3modesp}). Thus, OGLE-BLAP-148 is dominated by a transition from an earlier short-period $P_2$ state to a new longer-period $P_1$ state in DREAMS.

The light curves for OGLE-BLAP-148 are shown in Fig. \ref{fig:lc148100bin}. Similar to those for OGLE-BLAP-142, the pulsation amplitudes and shapes change over the years. The pulsation amplitude change from 0.15 mag (e.g., in 2017) to 0.25 (e.g., in 2019). The light curves appear to be more symmetric than those for OGLE-BLAP-142.

\subsection{Multiple Period Families and Mode Switching \label{sec:modes}}

\begin{figure*}
\centering
\includegraphics[width=\linewidth]{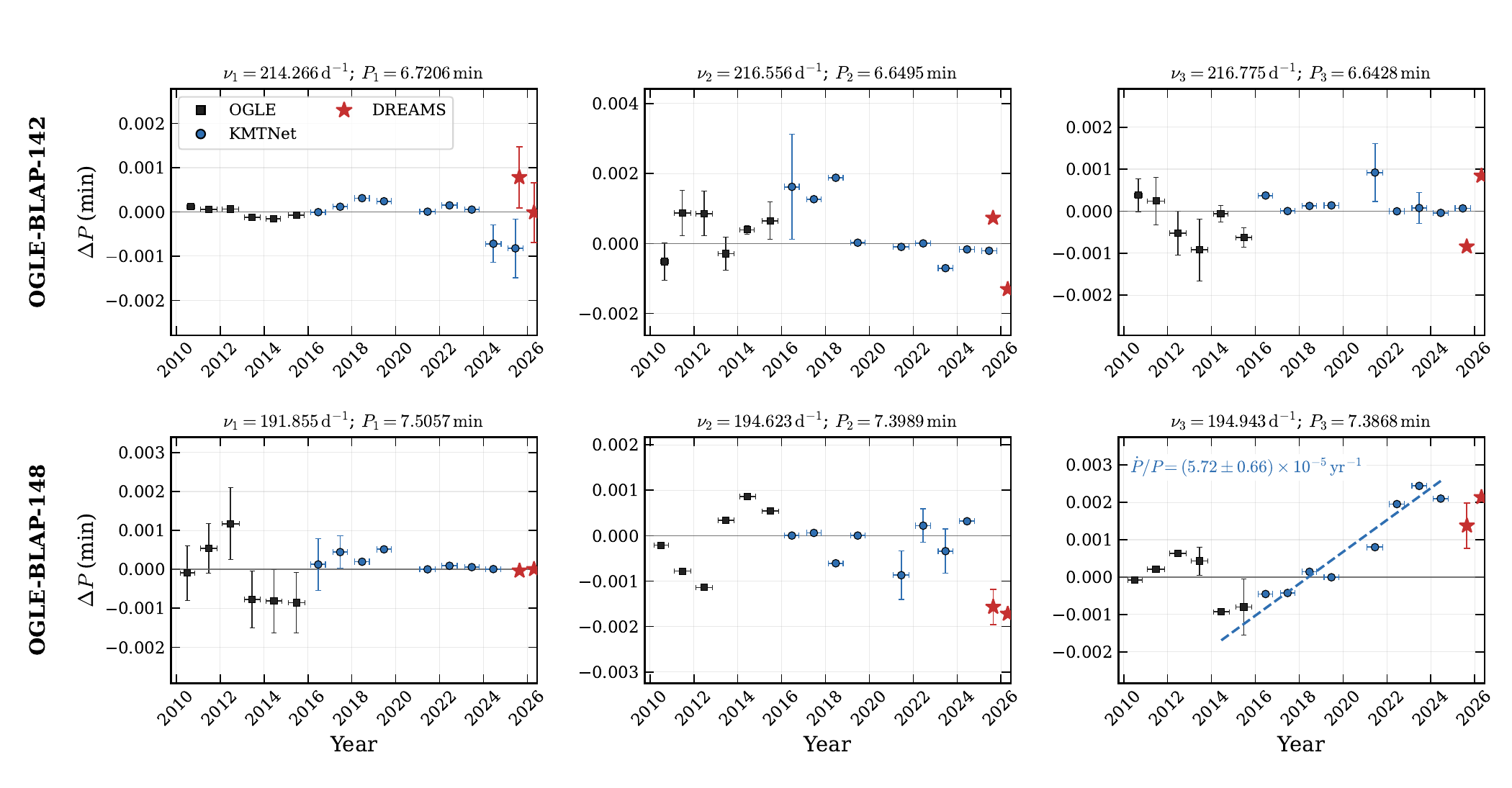}
\caption{Annual period offsets for the three candidate mode families in OGLE-BLAP-142 and OGLE-BLAP-148. Each panel shows $\Delta P = P_{\rm peak}-P_{\rm ref}$ relative to the reference frequency (\(P_{\rm ref}\)) labeled in the panel. Black squares, blue circles and red stars mark the OGLE, KMTNet and DREAMS periodogram peaks. The horizontal bar for each data point indicates the observing time span rather than an uncertainty while the vertical error bar shows the period uncertainty estimated from bootstrapping. The gray horizontal line denotes constant period. The dashed line in the lower middle panel shows the approximate secular drift of the \(P\approx7.3868\) min component in OGLE-BLAP-148 with $\dot{P}/P \approx 5.72 \times 10^{-5}\,{\rm yr}^{-1}$.}
\label{fig:3modesp}
\end{figure*}

Figure~\ref{fig:3modesp} shows the annual periods measured for the identified period families. The period uncertainty estimated from bootstrapping. In most panels, the yearly measurements do not follow a single smooth secular trend. Instead, they wander around the corresponding reference periods, with excursions whose formal year-to-year interpretation would imply apparent \(|\dot P/P|\) values ranging from \(\sim10^{-6}\) to \(\sim10^{-4}\,{\rm yr^{-1}}\). These large apparent rates should therefore not be interpreted automatically as evolutionary period changes; they more likely reflect changes in the relative amplitudes, phases, or detailed frequencies of the resolved pulsation families. The clearest exception is the intermediate-period family in OGLE-BLAP-148, shown in the lower-middle panel of Fig.~\ref{fig:3modesp}. This component, with \(P_3 \simeq 7.3868\,{\rm min}\), drifts almost linearly over 2014--2024. A linear fit over this interval gives $$\dot P/P = (5.72 \pm 0.66)\times10^{-5}\,{\rm yr^{-1}},$$ which uncertainty also estimated from bootstrapping. It is larger by a factor of \(\sim3\) than the largest linear \(\dot P/P\) value previously inferred for a BLAP \citep{2025AcA....75..223B}. However, the earlier 2010--2013 measurements and the latest 2025--2026 measurements do not lie on the extrapolation of this trend. We therefore interpret this linear segment as a coherent drift of one resolved period family over a limited time interval, rather than as evidence for a single long-term evolutionary \(\dot P\).

\begin{figure*}[t]
\centering
\includegraphics[width=0.9\textwidth]{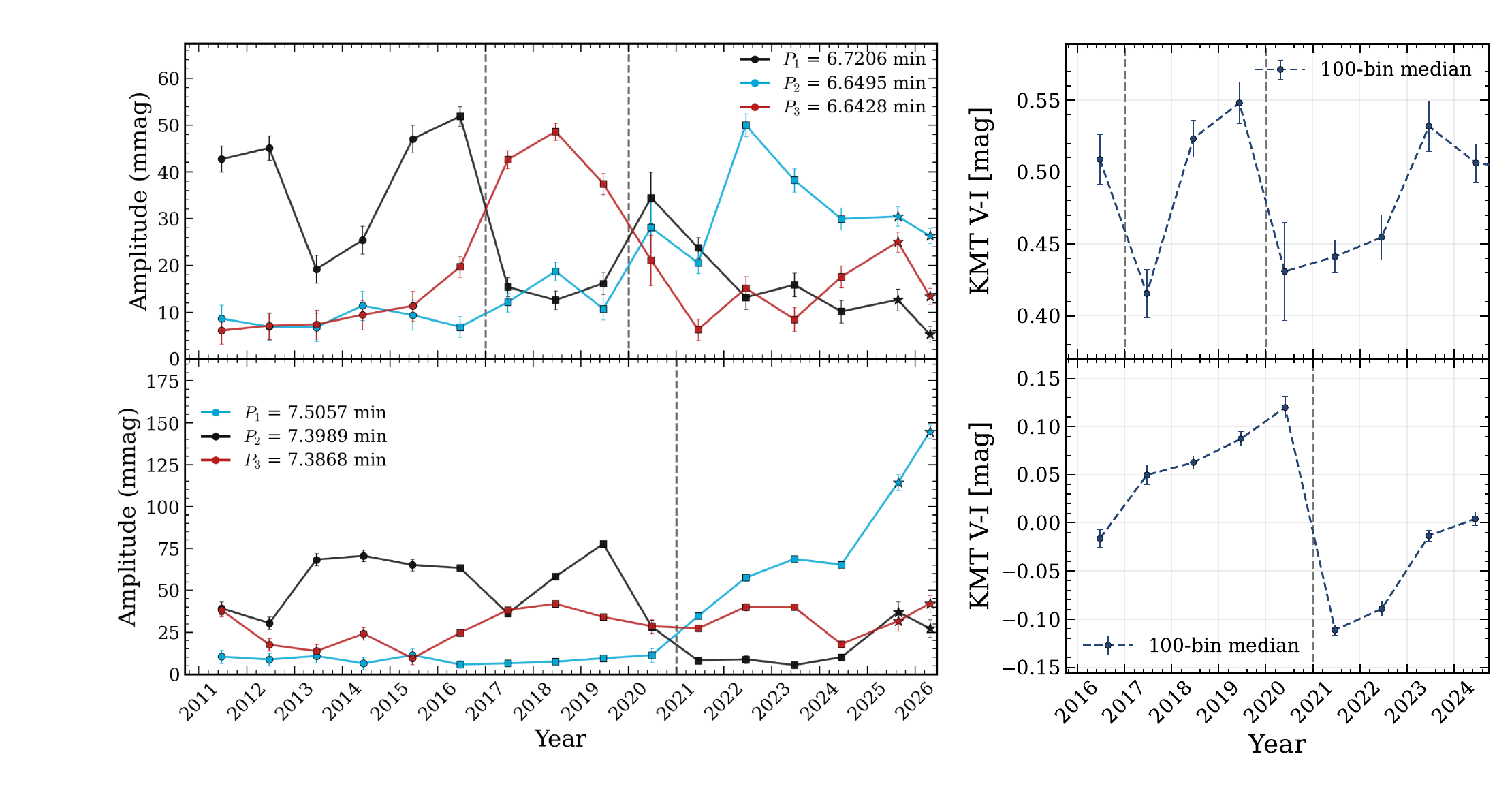}
\caption{Mode amplitudes and colors for OGLE-BLAP-142 (top) and OGLE-BLAP-148 (bottom) as a function of year. The left panels show the seasonal amplitudes of the selected mode families measured from OGLE (2011--2015, $I$ band), KMTNet (2016--2024, $I$ band), and DREAMS (2025--2026, $z$ band). The right panels show the corresponding KMTNet $V-I$ colors. The dashed lines indicates the timing of dominate mode shifts. Both stars appear to become bluer near the epochs when the dominant mode changes.}
\label{fig:amp142148}
\end{figure*}

The left two panels in Fig.~\ref{fig:amp142148} show the seasonal amplitudes. In OGLE-BLAP-142, the $P_1\approx6.7206$ min family weakens after the early seasons seen in OGLE, the $P_3\approx6.6428$ min family is strongest during the intermediate KMTNet seasons, while the $P_2\approx6.6495$ min family becomes dominant in the latest years. For OGLE-BLAP-148, the $P_2 \approx7.3989$ min family dominates before the transition, while the $P_1 \approx7.5057$ min family rises after 2021 and becomes strongest in DREAMS. These amplitude shifts clearly demonstrate the power struggle and transfer between these different modes.

The annual values in Fig.~\ref{fig:amp142148} are fitted single-sinusoid semi-amplitudes. We compare only the relative mode amplitudes within each season separately due to the difference of exposure times and band passes. Appendix~\S\ref{sec:amplitude_phase_sampling} tests whether such uneven phase coverage could bias the measured 2016--2024 KMTNet mode amplitudes, and we find that this has a negligible effect on the observed amplitude variations (by $\sim 0.6\%$).

The right panels of Fig.~\ref{fig:amp142148} show a possible color counterpart. For each KMTNet observing season, the calibrated $V$- and $I$-band light curves were phase-folded and partitioned into 100 phase bins. We calculated the median $V$ and $I$ magnitudes from the binned data and derived the corresponding $V-I$ color. The yearly color plotted in the figure is defined as the median of these phase-bin colors, with its uncertainty estimated from the median absolute deviation (MAD) of the binned V-I distribution. As mentioned in Section \ref{sec:dreams}, the color itself might not be accurate, but the year-to-year trends should be robust because they are insensitive to overall constant magnitude offsets.

In both stars the KMTNet $V-I$ color shifts to bluer color near the epochs when the dominant mode changes, with the clearest case being OGLE-BLAP-148 around 2021. Because the color measurements are sparse, we do not interpret them as precise temperature measurements. They nevertheless suggest that the mode exchange may be accompanied by a real change in the stellar atmosphere or envelope.

\section{Summary and Discussion \label{sec:discussion}}

Within the DREAMS footprint, we recovered all 14 BLAPs previously identified by OGLE and focused on OGLE-BLAP-142 and OGLE-BLAP-148, the two objects with the cleanest changes in their pulsation modes.

Our main result is that both stars change their dominant pulsation family between our observation, but the data do not firmly establish a characteristic switching or recurrence timescale. In OGLE-BLAP-142, the shortest securely traced dominant state is within the three-season interval from 2017 to 2019. In contrast, the 2020--2021 data constrained the modes poorly, while a different family clearly dominates from 2022 onward. OGLE-BLAP-148 shows only one transition, between the 2020 and 2021 seasons. To our knowledge, these are the two most secure identifications of year-scale mode switching in BLAPs. This result relies on the long-baseline combination of OGLE, KMTNet, and DREAMS: OGLE gives the early reference periods, KMTNet resolves the intermediate seasonal behavior, while DREAMS provides the highest-cadence and most recent epochs.

The detected period modes are separated by only about $\sim 1\%$. In OGLE-BLAP-142 the relevant periods are  $P_{1, 2, 3} \approx 6.7206$, $6.6495$, and $6.6428$ min, while for OGLE-BLAP-148 the main transition is between $P_2 \approx 7.3989$ and $P_1 \approx 7.5057$ min, with an additional mode near $P_3 \approx 7.3868$ min. The differences in period ratio are much smaller than that between the first overtone to the fundamental mode, $\approx 0.7-0.8$, as studied theoretically \citep{ByrneJeffery2020_FaintBlueStars,Jeffery2025_NonlinearBLAPModels,WuLi2025_OscillationPropertiesBLAPs,Das2026_InstabilityRegionsBLAPs}, and indeed seen in OGLE-BLAP-030, $\sim 0.762$ \citep{Pietrukowicz2025_ObservationalParameters}. Our observed mode switches are clearly not between these two modes. The exact mode identification is unclear.

The yearly period excursions correspond to apparent \(|\dot P/P|\) values of order \(10^{-6}\)--\(10^{-4}\,{\rm yr^{-1}}\). While most of the identified mode frequencies seem to oscillate, one component in OGLE-BLAP-148 drifts linearly for $\sim$ a decade (see the middle panel of Fig.～\ref{fig:3modesp}). The $P\approx7.3868$ min component ($P_3$) has \(\dot P/P\approx 5.72\times10^{-5}\), more than a factor of three larger than the $\approx 1.6\times 10^{-5}\,{\rm yr}^{-1}$ values reported for OGLE-BLAP-153 and OGLE-BLAP-176 \citep{2025AcA....75..223B}. However, this drift behavior does not persist in the earliest (2010-2013) and latest data (2025-2026).

Multi-mode behavior may be particularly common at the short-period end of the BLAP population. Among the 14 OGLE BLAPs overlapping with DREAMS, three have periods shorter than 10 min, and two of them (studied in detail here), OGLE-BLAP-142 and OGLE-BLAP-148, show clear multi-periodic structure and dominant-mode switching. This small-number statistics should not be interpreted as an occurrence rate, but it strongly suggests that the short-period regime may be especially favorable for exciting several closely spaced pulsation families. The short-period BLAPs are likely to overlap strongly with the high-gravity BLAP population \citep{Kupfer2019_HighGravityBLAPs,McWhirter2020_MonoperiodicityBLAPs}, nevertheless spectroscopic constraints on \(T_{\rm eff}\) and especially \(\log g\) are needed to verify whether OGLE-BLAP-142 and OGLE-BLAP-148 indeed belong to this population. If confirmed, the high incidence of multi-periodicity among these short-period objects would provide new constraints on the differences between high-gravity BLAPs and their longer-period counterparts \citep{Byrne2021_BinaryEvolutionBLAPs,Xiong2022_ShellHeBurningBLAPs, Lin2023_TMTSBLAP1,Koen2024_ZTFJ071329}.

The color evolution may provide an additional clue. The right panels of Fig.~\ref{fig:amp142148} show that both objects shift to bluer colors, i.e. smaller \(V-I\), near the epochs when the dominant period family changes. This may indicate a change in the phase-averaged \(T_{\rm eff}\), or a different temperature response of the newly dominant mode. However, broad-band colors alone are not unique diagnostics in crowded bulge fields, because they can be affected by phase sampling, color-dependent pulsation amplitudes, blending, and calibration systematics. Since phase-resolved spectroscopy of BLAPs has shown that \(T_{\rm eff}\), \(\log g\), and radial velocity can vary substantially over the pulsation cycle \citep{Bradshaw2024_OGLEBLAP009}, spectroscopy is needed to test whether the color shifts reflect real atmospheric changes associated with the mode switch.

The physical origin of these multiple frequencies and their switching mechanism remains unclear. The observed mode switching is unlikely to be caused by shell flashes, or thermal-pulses as these violent dynamic events occur rarely \citep{Iben1983FinalFlash,Asplund1999Sakurai,Herwig2001SakuraiTimescale,Hajduk2005Sakurai,MillerBertolami2007VLTP,Lau2011V605Aql,Clayton2013V605Aql,Reindl2017SAO244567,Driebe1999HeWDFlashes,Althaus2001DiffusionHFlashes,Istrate2014ProtoHeWD,Istrate2016LowMassHeWD}. 

The central issue is why several closely spaced modes (within $\sim$ 1\%) are excited and why their relative amplitudes can switch within a timescale of $\sim$ year. Mode appearance and disappearance are often seen in white dwarfs \citep{Kleinman1998, Uzundag2023, BischoffKim2019} over $\sim$ decades. However, real white dwarf period instabilities are slow, surface/rotation-driven (see, e.g., \citealt{WuGoldreich2001, Wu2001}), and orders of magnitude smaller in terms of $\dot{P}/P$ than in BLAPs. In particular, for white dwarfs, new frequencies may (dis)appear, but they frequencies are often not fixed. In comparison,  for the two BLAPs studied here, the three frequencies are stable, but their relative amplitudes change. We also note that amplitude modulation via nonlinear mode interactions is relatively common among classical pulsators: an ensemble study of 983 Kepler δ Sct stars found that $\sim 60\%$ exhibit significant amplitude variability, attributable to mechanisms including beating and genuine nonlinear resonant coupling \citep{Bowman2016}. However, these are pulsations for more massive stars ($\sim 2 M_\odot$). More tailored theoretical studies (e.g., along the lines of \citealt{Jeffery2025_NonlinearBLAPModels, WuLi2025_OscillationPropertiesBLAPs}) for BLAPs which have lower mass are needed to understand these distinct observed properties and explain their high occurrence rate at the short-period regime.

Observationally, the next step is low-resolution optical spectroscopy on an 8--10 m class telescope, such as Gemini-S/GMOS-S \citep{Hook2004_GMOS}. The main goal is to measure \(T_{\rm eff}\) and \(\log g\) from the Balmer and, where present, helium line profiles, and thereby test whether these short-period objects belong to the high-gravity BLAP population \citep{Kupfer2019_HighGravityBLAPs,McWhirter2020_MonoperiodicityBLAPs}. High spectral resolution is not required for this first step: for OGLE-BLAP-009, \citet{Bradshaw2024_OGLEBLAP009} measured \(T_{\rm eff}\) and \(\log g\) using low-resolution Keck/LRIS spectra with \(R\simeq1000\). However, OGLE-BLAP-142 and OGLE-BLAP-148 are about \(4\) mag fainter in the \(I\) band than OGLE-BLAP-009, implying roughly \(40\)--\(60\) times lower flux. Because the switching is currently resolved on year timescales, one spectroscopic epoch per observing season represents the minimum useful cadence. If future DREAMS data localize the switching to shorter timescales, e.g., $\sim 3$ months, the cadence should therefore be increased accordingly.
 
\begin{acknowledgments}

We thank Yanqin Wu, Xuefei Chen and Yaguang Li for helpful discussions, and the referee for a constructive report that improved the paper.
H.M., S.M., H.Y., W.Z, Z.W., Q.Q., Y.T. and Y.S. acknowledge support from the National Natural Science Foundation of China (Grant No. 12133005; PI: S.M.). H.Y. acknowledges support from the China Postdoctoral Science Foundation (No. 2024M762938). This research has made use of the KMTNet system operated by the Korea Astronomy and Space Science Institute (KASI) at three host sites of CTIO in Chile, SAAO in South Africa, and SSO in Australia.
Data transfer from the host site to KASI was supported by the Korea Research Environment Open NETwork (KREONET). The OGLE project has received funding from the Polish National Science Centre grant OPUS 2024/55/B/ST9/00447 awarded to A.U. The authors thank the High Performance Computing Center at Westlake University for providing computational and data-storage resources, and the Office of Information Technology at Westlake University for assistance with data transfer and with the development and deployment of the data-release site. This research was funded in part by National Science Centre, Poland, grant SONATA 2023/51/D/ST9/00187 awarded to P.M. 

T. Wu thanks the co-supports from the National Natural Science Foundation of China (Grant No. 12288102), from B-type Strategic Priority Program of the Chinese Academy of Sciences (Grant No. XDB1160202), and from the National Key R\&D Program of China  (Grant No. 2021YFA1600400/2021YFA1600402).  TW also gratefully acknowledge the supports of NSFC of China (Grant Nos. 12133011 and 12273104), Yunnan Fundamental Research Projects (Grant No. 202401AS070045), and Yunnan Key Laboratory of Supernova Research (No. 202505AV340004).

This project used data obtained with the Dark Energy Camera (DECam), which was constructed by the Dark Energy Survey (DES) collaboration. Funding for the DES Projects has been provided by the U.S. Department of Energy, the U.S. National Science Foundation, the Ministry of Science and Education of Spain, the Science and Technology Facilities Council of the United Kingdom, the Higher Education Funding Council for England, the National Center for Supercomputing Applications at the University of Illinois at Urbana-Champaign, the Kavli Institute for Cosmological Physics at the University of Chicago, the Center for Cosmology and Astro-Particle Physics at The Ohio State University, the Mitchell Institute for Fundamental Physics and Astronomy at Texas A\&M University, Financiadora de Estudos e Projetos, Funda\c{c}\~ao Carlos Chagas Filho de Amparo \`a Pesquisa do Estado do Rio de Janeiro, Conselho Nacional de Desenvolvimento Cient\'{\i}fico e Tecnol\'ogico and the Minist\'erio da Ci\^encia, Tecnologia e Inova\c{c}\~ao, the Deutsche Forschungsgemeinschaft, and the collaborating institutions in the Dark Energy Survey.

The collaborating institutions are Argonne National Laboratory; the University of California at Santa Cruz; the University of Cambridge; Centro de Investigaciones Energ\'eticas, Medioambientalesy Tecnol\'ogicas (CIEMAT), Madrid; the University of Chicago; University College London; the DES-Brazil Consortium; the University of Edinburgh; the Eidgen\"ossische Technische Hochschule (ETH) Z\"urich; Fermi National Accelerator Laboratory; the University of Illinois at Urbana-Champaign; the Institut de Ci\`encies de l'Espai (IEEC/CSIC); the Institut de F\'{\i}sica d'Altes Energies (IFAE); Lawrence Berkeley National Laboratory; the Ludwig-Maximilians-Universit\"at M\"unchen and the associated Excellence Cluster Universe; the University of Michigan; NSF NOIRLab; the University of Nottingham; The Ohio State University; the OzDES Membership Consortium; the University of Pennsylvania; the University of Portsmouth; SLAC National Accelerator Laboratory; Stanford University; the University of Sussex; and Texas A\&M University.

This project used data from the DECam Rogue Earths and Mars Survey (DREAMS), whose primary light-curve generation and archive are hosted by the Department of Astronomy at Westlake University. Based on observations at NSF Cerro Tololo Inter-American Observatory, NSF NOIRLab (NOIRLab Prop.\ ID 2025A-806294, PI: Weicheng Zang; Prop.\ ID 2025B-560332, PI: Weicheng Zang \& Hongjing Yang), which is managed by the Association of Universities for Research in Astronomy (AURA) under a cooperative agreement with the U.S. National Science Foundation.

\end{acknowledgments}

\facilities{CTIO: Blanco (DECam); KMTNet; OGLE}
\software{NumPy \citep{numpy:2020}, 
          Matplotlib \citep{Matplotlib}, 
          SciPy \citep{scipy:2020}, 
          H5py \citep{h5py},
          Astropy \citep{astropy}}

‌\bibliographystyle{aasjournalv7}
‌\bibliography{reference}

\newpage

\restartappendixnumbering
\appendix

\section{Folded light curves}
\label{sec:folded-lightcurves}

\begin{figure*}
\centering
\includegraphics[width=0.80\linewidth]{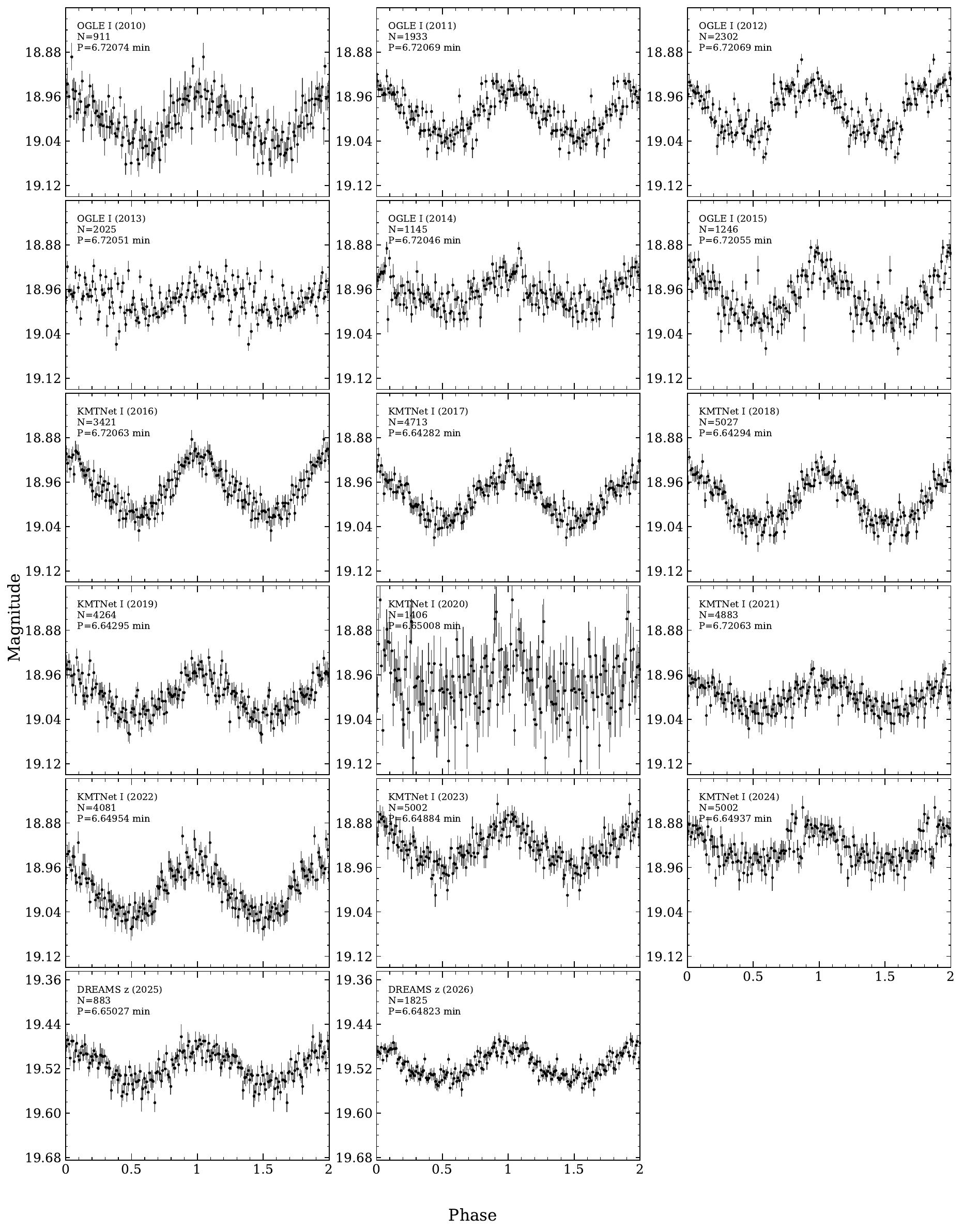}
\caption{Folded light curves for OGLE-BLAP-142 as a function of season. OGLE and KMTNet points are in the $I$ band, while DREAMS points are in the $z$ band. The adopted folding period for each year is labeled at the top. Notice that the zero-point, pulsation amplitude and light-curve shape change with time. $N$ gives the number of data points in each season. For clarity, the phase-folded data points are grouped into 100 equally spaced bins across the entire phase.}
\label{fig:lc142100bin}
\end{figure*}

\begin{figure*}
\centering
\includegraphics[width=0.80\linewidth]{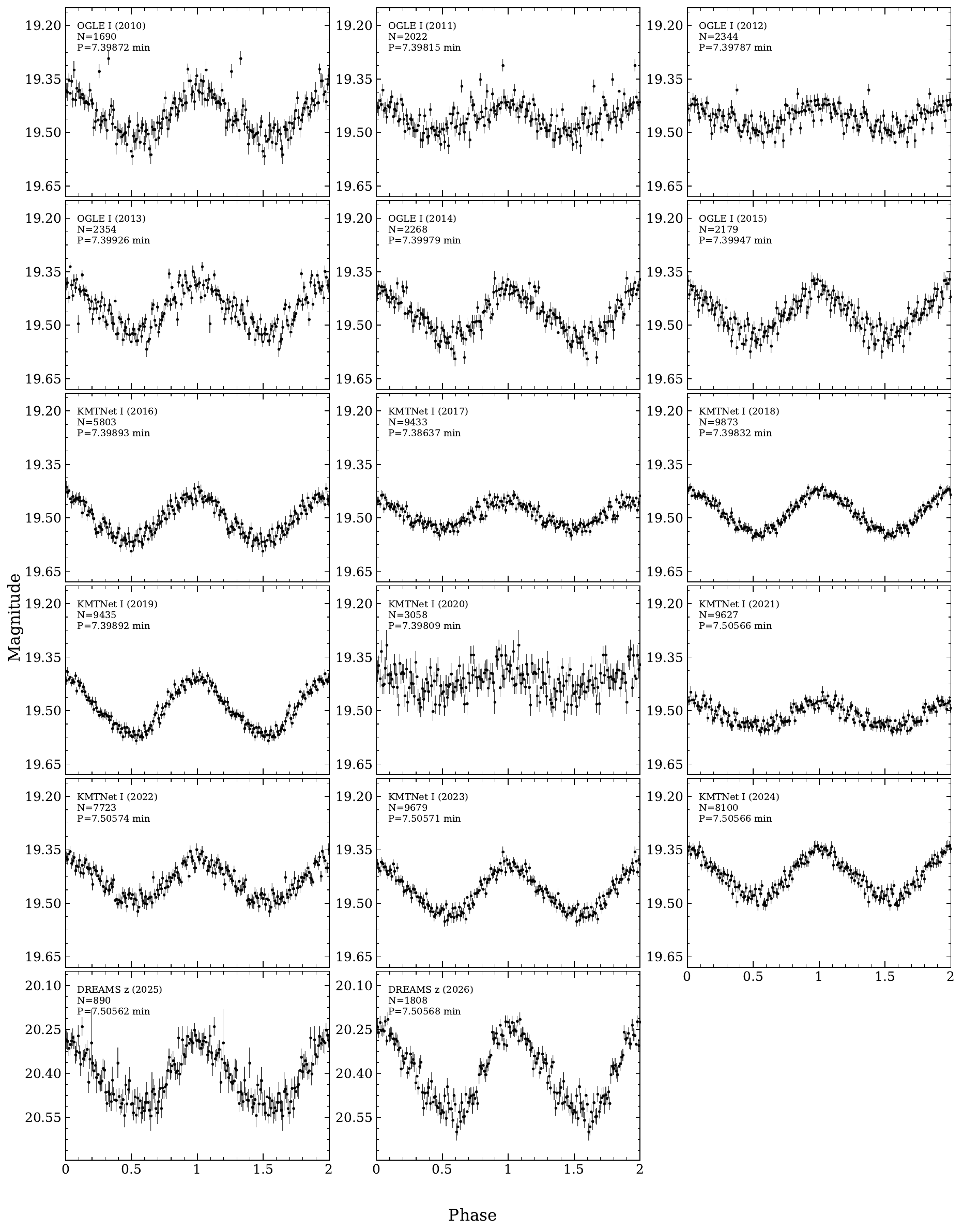}
\caption{Folded light curves for OGLE-BLAP-148 as a function of season. Notation follows Fig.~\ref{fig:lc142100bin}.}
\label{fig:lc148100bin}
\end{figure*}

\clearpage

In this section, we show the light curves for OGLE-BLAP-142 and OGLE-BLAP-148 in Figs.\,\ref{fig:lc142100bin} and \ref{fig:lc148100bin}, respectively.

\section{Amplitude Attenuation Due to Finite exposure}
\label{sec:observing_setup}

The period and amplitude analyses use 100~s OGLE $I$-band exposures, mainly 60\,s KMTNet $I$-band exposures (with a small number of data points having exposure times between 10 and 90\,s), and DREAMS $z$-band exposures of 42\,s in 2025A and 2026 and 60\,s in 2025B. OGLE $V$-band exposures are 150\,s and provide catalog means only, while the 90\,s KMTNet $V$-band exposures are used only for the color analysis (see Fig.~\ref{fig:amp142148}).

For a sinusoidal signal of period \(P\), a boxcar integration of duration \(t_{\rm exp}\), timestamped at its midpoint, multiplies the measured semi-amplitude by
\begin{equation}
\eta(t_{\rm exp},P)=\frac{\sin(\pi t_{\rm exp}/P)}{\pi t_{\rm exp}/P}
\label{eq:ita}
\end{equation}
which is introduced in Eq.~2 of ~\citet{2014MNRAS.445..946C}. Across the six candidate periods (3 for each BLAP), 6.6428--7.5057~min, we found \(\eta=0.900\)--0.921 for OGLE $I$, 0.963--0.971 for the KMTNet $I$ and 60~s DREAMS exposures, and 0.982--0.986 for 42~s DREAMS exposures. Thus, for the exposure times used here, finite integration reduces the measured semi-amplitudes by only from 1.4\% (DREAMS) to 10.0\% (OGLE). The key seasonal mode-amplitude changes, however, are measured within the KMTNet $I$-band data during which all modes share the same exposures. For the typical 60\,s exposure, $t_{\rm exp}/P\simeq0.14$, and a Taylor expansion of Eq.~(\ref{eq:ita}) gives
\begin{equation}
\eta \simeq 1-\frac{1}{6}\left(\frac{\pi t_{\rm exp}}{P}\right)^2.
\label{eq:taylor}
\end{equation}
At fixed exposure time, the difference in attenuation between modes is

\begin{equation}
|\Delta\eta|\simeq \left|\frac{d\eta}{dP}\Delta P\right|\simeq \frac{1}{3}\left(\frac{\pi t_{\rm exp}}{P}\right)^2\frac{|\Delta P|}{P}.
\label{eq:sim}
\end{equation}
Since the mode periods differ by at most about $1.6\%$, this gives $|\Delta\eta|\sim10^{-3}$. Even for the longest 90\,s KMTNet exposures, $\Delta\eta$ remains below 0.25\%. Thus, finite exposure times has no measurable effect on the periods or on which mode has the largest amplitude.

\section{Effects of irregular Nyquist Frequency on mode identification}
\label{sec:sampling_validation}

Our data do not have a unique classical Nyquist frequency determined by a single sampling interval because the OGLE and KMTNet sampling is non-uniform; in addition, KMTNet combines observations from three sites \citep{VanderPlas2018_LombScargle}. For illustration only, The median inter-exposure intervals ($\Delta t$)  gives formal \(f_{\rm Nyq}=1/(2\Delta t)\) ranges of \(31.3\)--\(36.2~{\rm day}^{-1}\) for OGLE, \(26.8\)--\(84.9~{\rm day}^{-1}\) for KMTNet, and \(598\)--\(601~{\rm day}^{-1}\) for DREAMS. Thus, the \(191.9\)--\(216.8~{\rm day}^{-1}\) pulsations are formally above a median-cadence Nyquist value for OGLE and KMTNet, but this regular-sampling expression is not an alias-free upper-frequency limit for these data. Below we will test the recoverability directly from the actual sampling.

For the three  modes for each star in every seasons separately, we first fitted and removed all three modes and their first harmonics from the observed light curves. To generate noise, we divided the residuals by their photometric errors, randomly resampled these values with replacement, and multiplied them by the original errors. We then added a sinusoid at the actual observation times, using the fitted frequency \(f_{\rm inj}\), semi-amplitude \(A_{\rm inj}\), and a random phase. We generated 500 such light curves for each mode and analyzed them with the same GLS search, recording the frequency \(f_{\rm rec}\) of the highest peak.

\begin{figure*}
\centering
\includegraphics[width=\linewidth]{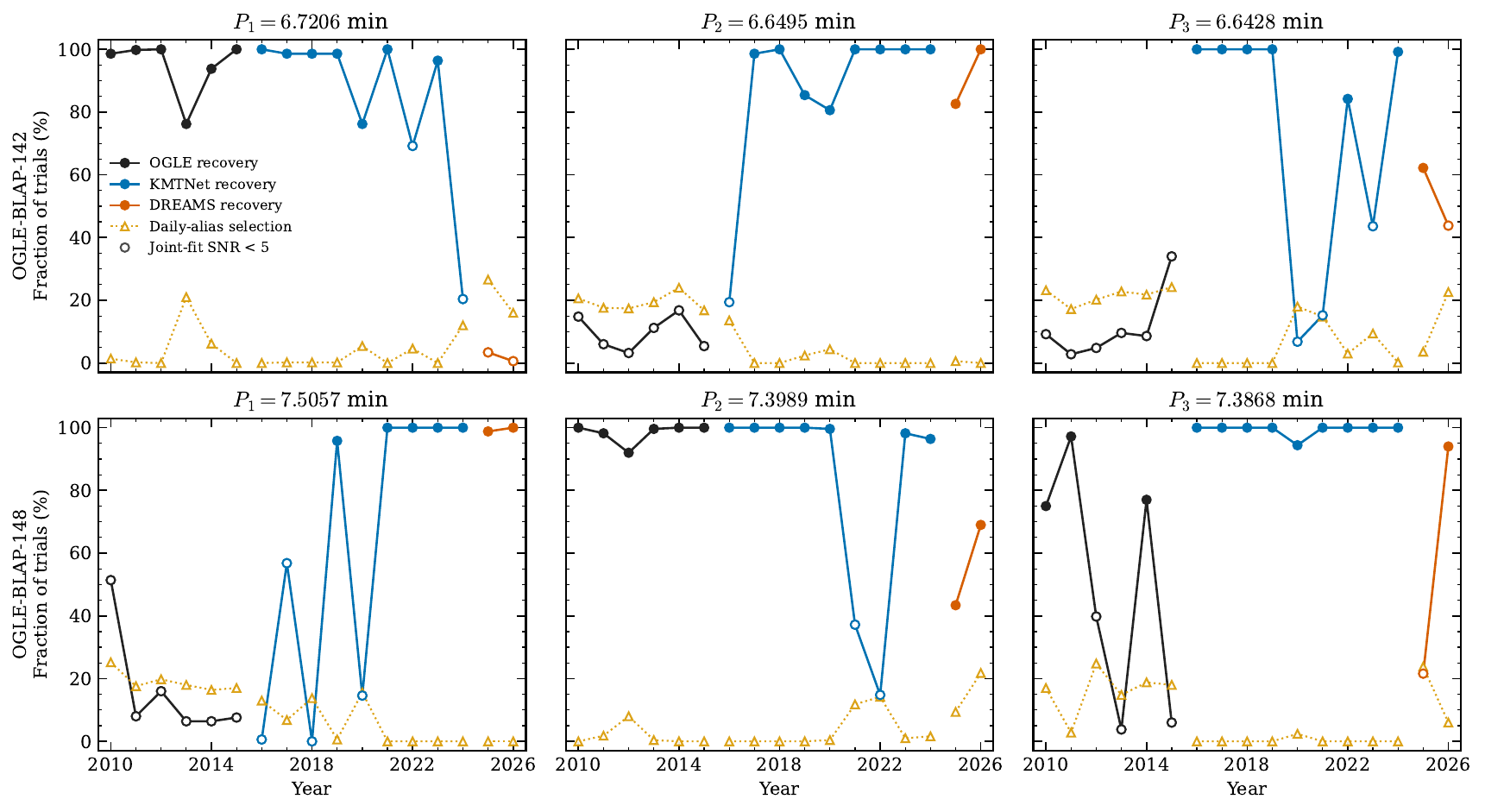}
\caption{Injection--recovery results for OGLE-BLAP-142 (top) and OGLE-BLAP-148 (bottom), with one candidate mode per column. Black, blue, and red solid curves show the strict-recovery fractions for OGLE, KMTNet, and DREAMS, respectively. Pale-orange dotted curves with open triangles show the integer-day alias fractions. Filled circles show those with \(A/\sigma_A\geq5\) where \(\sigma_A\) is the semi-amplitude uncertainty from the simultaneous fit of all three modes and their first harmonics. These high S/N ratio modes are recovered with in most cases (high fraction). Open circles indicates modes with \(A/\sigma_A<5\), and they are less well recovered due to lower S/N ratios. Open triangles show the fractions where daily-alias frequencies have been selected as the dominant ones. They are quite low, indicating in most cases the injected modes are not mis-identified as daily-aliases.}
\label{fig:sampling_validation}
\end{figure*}

We counted a trial as a strict recovery when \(\lvert f_{\rm rec}-f_{\rm inj}\rvert\leq1/T\), where \(T\) is the observing time span of that observing season. We also used a wider tolerance of \(0.08~{\rm day}^{-1}\) to distinguish peaks near the injected frequency from its daily aliases. Peaks within this tolerance of \(f_{\rm inj}\) were assigned to the injected mode; those near \(f_{\rm inj}\pm n~{\rm day}^{-1}\), with \(n=1,2,\ldots\), were counted as daily aliases. Figure~\ref{fig:sampling_validation} shows the fraction of trials that recovered the injected frequency under the strict constraint and the fraction that was mis-identified as a daily alias, indicating that the dominant seasonal family is well recovered from the actual timestamps.

\section{Effect of Uneven Phase Sampling on Seasonal Amplitudes}
\label{sec:amplitude_phase_sampling}

Uneven phase sampling may affect amplitude measurements by giving unbalanced weight to certain parts of the pulsation period \citep{Lanzafame2018}. For example, if all our observations happen to fall within only part of the descending branch, we may underestimate the amplitude. By using the 2016--2024 KMTNet $I$-band data, we test whether giving each phase bin equal total weight changes the fitted amplitudes.
 
Adjusting the weights according to phase coverage can reduce the influence of densely sampled regions \citep{Rimoldini2014}. The original semi-amplitudes in Fig.~\ref{fig:amp142148}, denoted by $A_{\rm original}$ below, were obtained using inverse-variance weights $w_i=\sigma_i^{-2}$. At the adopted seasonal frequency $f$, the pulsation phase of an observation at time $t_i$ is
\begin{equation}
\phi_i=(t_i-t_0)\bmod P, ~~~ {P=\frac{1}{f}}
\label{eq:phase_sampling_phase}
\end{equation}
where $t_0$ is an epoch of the maximum magnitude of the fitted sinusoid. To test the effect of unequal phase weighting, we grouped the observations into 20 equal phase bins and rescaled their weights. The original inverse-variance weights, $w_i=\sigma_i^{-2}$, give the $k$-th bin a total weight $W_k=\sum_{j \in \{B_k\}}w_j$, where $\{B_k\}$ selects all the data points that belong to the $k$-th bin. To remove differences in weights between bins, we assigned each observation a rescaled weight
\begin{equation}
\widetilde{w}_i=\frac{w_i}{\displaystyle\sum_{j\in B_{k}}w_j}.
\label{eq:phase_balanced_weights}
\end{equation}
This normalization gives every phase bin the same total weight and preserving the relative inverse-variance weights within each bin. We then repeated the fit at fixed $f$ using $\widetilde{w}_i$ to obtain $A_{\rm rescaled}$.

\begin{figure*}
\centering
\includegraphics[width=\linewidth]
{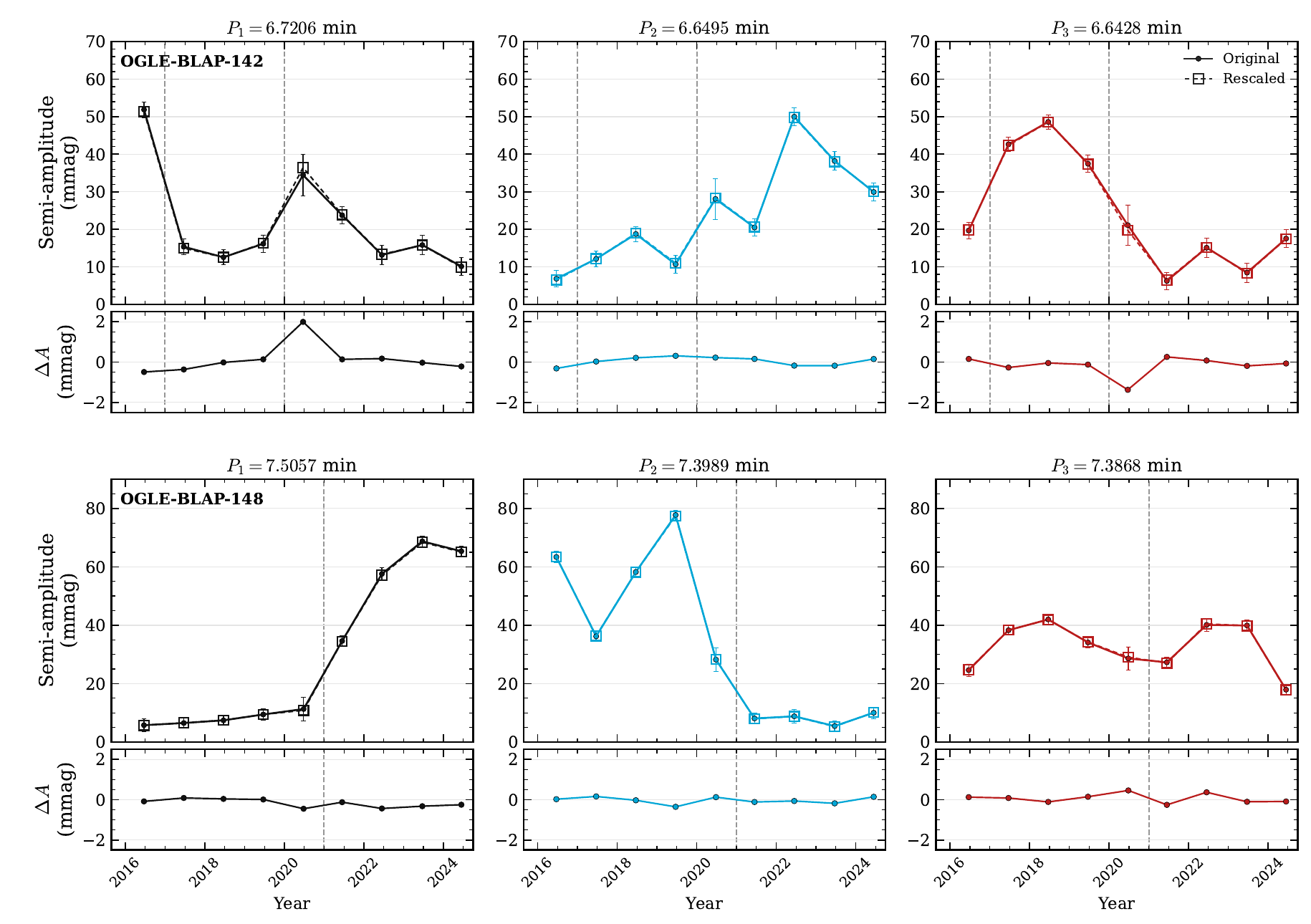}
\caption{Original inverse-variance-weighted KMTNet semi-amplitudes $A_{\rm original}$ (filled circles) and phase-rescaled amplitudes $A_{\rm rescaled}$ (open squares) for OGLE-BLAP-142 (top) and OGLE-BLAP-148 (bottom).The mode labels $P_1$, $P_2$, and $P_3$ follow the definitions in the main text, with the corresponding curves shown in black, blue, and red. The lower panels show $\Delta A=A_{\rm rescaled}-A_{\rm original}$, which indicates the two estimates are nearly indistinguishable. The vertical dashed lines mark the mode-transition epochs in Fig.~\ref{fig:amp142148}.}
\label{fig:phase_balanced_amplitudes}
\end{figure*}

The median value of $\lvert A_{\rm rescaled}-A_{\rm original}\rvert$ was only $0.15$~mmag, and the overall maximum was $2.00$~mmag in the sparsely sampled 2020 season of OGLE-BLAP-142 due to the relatively poor quality data. For other seasons，the effect is even smaller; for example, the largest change in 2019 data was only $0.35$~mmag. Repeating the analysis with 10 and 40 bins gave median changes of $0.09$ and $0.18$~mmag, respectively. These changes are negligible compared with the observed seasonal amplitude ranges of $24.1$--$72.3$~mmag (Fig.~\ref{fig:phase_balanced_amplitudes}). These results indicate that the observed irregularity in phase coverage has little effect on our seasonal amplitude measurements.

\end{CJK*}
\end{document}